\documentclass[twocolumn]{aastex631}

\usepackage{multirow}
\usepackage{natbib}

\def\h{\hskip 0.0 mm}
\def\nd{\nodata}

\def\apjl{{ApJL}}
\def\apjs{{ApJS}}
\def\asec{$^{\prime\prime}$}

\def\e#1{$\times$10$^{#1}$}

\def\farcs{\hbox{$.\mkern-4mu^{\prime\prime}$}}

\def\hal{H$\alpha$}
\def\hb{H$\beta$}
\def\pal{Pa$\alpha$}
\def\pb{Pa$\beta$}

\def\kms{km s$^{-1}$}

\def\lax{{$\mathrel{\hbox{\rlap{\hbox{\lower4pt\hbox{$\sim$}}}\hbox{$<$}}}$}}
\def\gax{{$\mathrel{\hbox{\rlap{\hbox{\lower4pt\hbox{$\sim$}}}\hbox{$>$}}}$}}
\def\simlt{\lower.5ex\hbox{$\; \buildrel < \over \sim \;$}}
\def\simgt{\lower.5ex\hbox{$\; \buildrel > \over \sim \;$}}
\def\lum{erg s$^{-1}$}

\def\cm2{cm$^{-2}$}

\def\heii{\ion{He}{2}}

\def\oiii{[\ion{O}{3}]}

\def\nii{[\ion{N}{2}]}

\def\sii{[\ion{S}{2}]}

\def\mm{$M_{\rm BH}-M_{\star}$}

\def\msig{$M_{\rm BH}-\sigma_\star$}

\def\ser{S\'{e}rsic}

\shorttitle{MIR flare in NGC 3786}
\shortauthors{Son et al.}

\begin{document}

\title{A Mid-infrared Flare in the Seyfert Galaxy NGC 3786: A Changing-look Event Triggered by an Obscured Tidal Disruption Event?}

\author[0000-0002-5346-0567]{Suyeon Son}
\affiliation{Department of Astronomy and Atmospheric Sciences,
Kyungpook National University, Daegu 41566, Korea; mkim.astro@gmail.com}

\author[0000-0002-3560-0781]{Minjin Kim}
\affiliation{Department of Astronomy and Atmospheric Sciences, 
Kyungpook National University, Daegu 41566, Korea; mkim.astro@gmail.com}

\author[0000-0001-6947-5846]{Luis C. Ho}
\affiliation{Kavli Institute for Astronomy and Astrophysics, Peking 
University, Beijing 100871, China}
\affiliation{Department of Astronomy, School of Physics, Peking University, Beijing 100871, China}

\author[0000-0002-6925-4821]{Dohyeong Kim}
\affiliation{Department of Earth Sciences, Pusan National University, Busan 46241, Korea}

\author[0000-0002-5857-5136]{Taehyun Kim}
\affiliation{Department of Astronomy and Atmospheric Sciences, 
Kyungpook National University, Daegu 41566, Korea; mkim.astro@gmail.com}

\correspondingauthor{Minjin Kim}
\email{mkim.astro@gmail.com}

\begin{abstract}
We report an exceptional mid-infrared flare in the Seyfert 1.8 NGC 3786. In the multi-epoch data from the Wide-field Infrared Survey Explorer, the nuclear mid-infrared brightness of NGC 3786 appears to vary substantially up to $0.5-0.8$ mag around mid-2020. However, there is no evidence of significant variation in the corresponding light curve of the optical band from the Zwicky Transient Facility. This implies that the flare may have been heavily obscured by nuclear dust. Through follow-up spectroscopic observations with Gemini-North after the flare, we find that broad emission lines in \pal\ and \pb\ newly appear, while the broad \hb\ emission is marginally detected in the post-flare spectrum. In addition, their central wavelengths are systematically redshifted up to 900 \kms\ with respect to the narrow emission lines. This reveals that the flare is associated with the changing-look phenomenon from type 1.8 to type 1. Based on these findings, we argue that the flare is likely to originate from an obscured tidal disruption event, although extreme variation in the accretion rate may not be ruled out completely.
    
\end{abstract}


\section{Introduction} 

Supermassive black holes (SMBHs) are ubiquitous, at least at the center of massive galaxies (e.g., \citealt{kormendy_2013}). A majority of SMBHs remain dormant. If a wandering star approaches an SMBH sufficiently close to be tidally disrupted, an accretion disk with a super-Eddington ratio is suddenly formed from the tidal debris of the star and a large amount of energy is emitted through X-rays and ultraviolet (UV). This phenomenon is known as a tidal disruption event (TDE; \citealt{hills_1975, rees_1988}). TDEs can prove the demography of SMBHs in quiescent galaxies and the physical properties of central stars in galactic nuclei (e.g., \citealt{stone_2016, graur_2018, french_2020}).

To date, a few tens of TDEs have been observed using multi-epoch data in the X-ray and UV/optical bands (e.g., \citealt{komossa_1999, gezari_2006, velzen_2021, gezari_2021}). Although observed TDE rates appear to be lower than those that are theoretically predicted, the origin of this discrepancy is unknown (e.g., \citealt{magorrian_1999, holoien_2016}). One possibility is that TDEs that occur in the obscured nucleus may be difficult to discover in the existing dataset using current detection methods (e.g., \citealt{tadhunter_2017, kool_2020}). Therefore, to examine TDE phenomena fully, it is essential to investigate the occurrence rate and physical properties of obscured TDEs.

Similar to an active galactic nucleus (AGN), a TDE can exhibit IR emission radiated from the circumnuclear dust heated by the X-ray/UV continuum from the accretion disk (e.g., \citealt{komossa_2009, lu_2016, velzen_2016, jiang_2016}). Such IR echoes in optically selected TDEs have been studied systematically using the light curves of the Wide-field Infrared Survey Explorer (WISE; e.g., \citealt{jiang_2021}). \citet{jiang_2021} argued that the dust covering factor is less than 0.01 and reported that known TDEs are substantially biased toward those which are minimally obscured.

\begin{figure}[t!]
\centering
\includegraphics[width=0.5\textwidth]{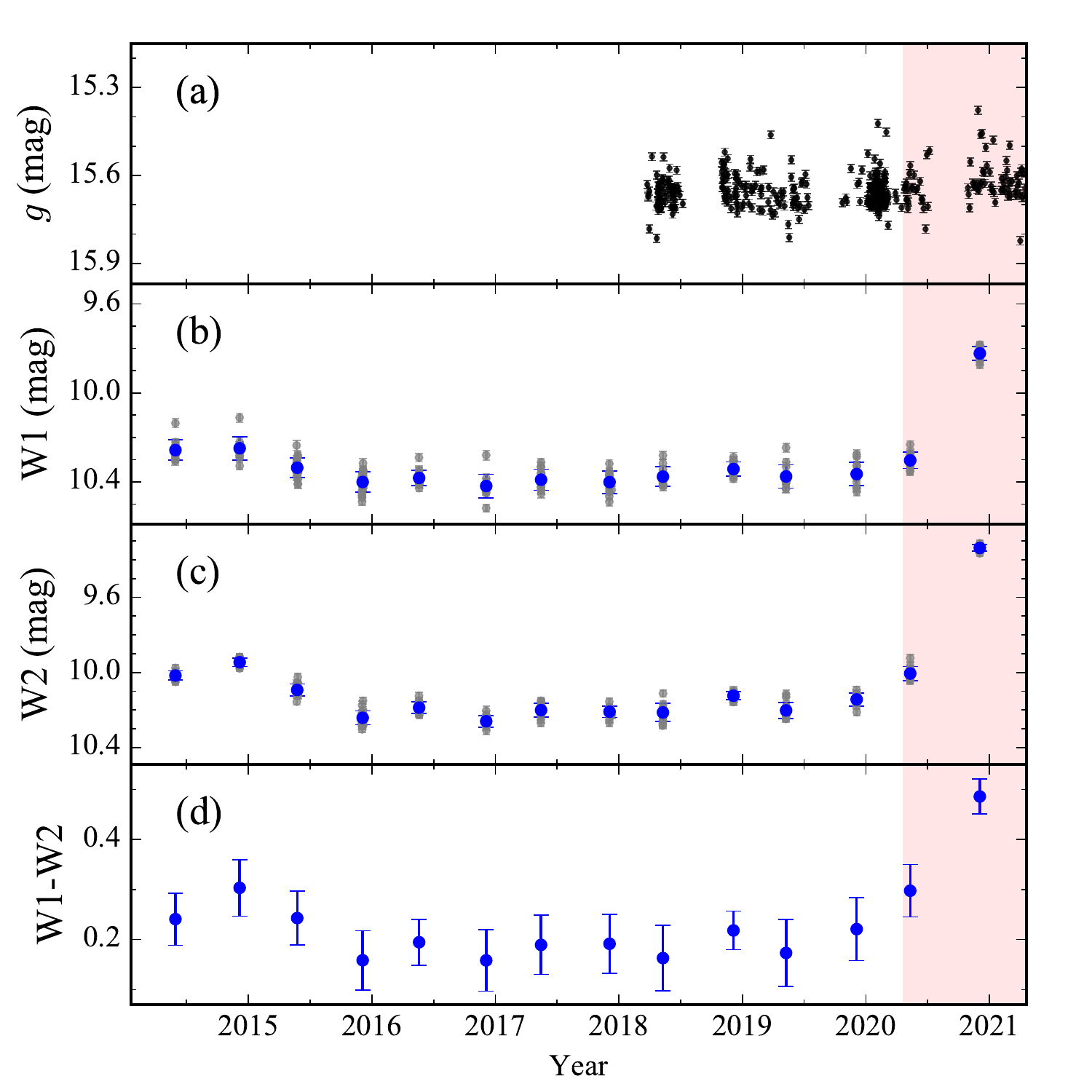}
\caption{
Light curves of NGC 3786 in the (a) $g$ band from the Zwicky Transient Facility, (b) WISE W1 band, (c) WISE W2 band, and (d) W1-W2 color.  In panels (b) and (c), the gray circles denote photometric data from individual visits of WISE, and blue circles represent the median values in each epoch. The error bars represent the $1\sigma$ uncertainty. The epoch of the MIR flare is highlighted by the red shaded area.
\label{fig:fig1}}
\end{figure}

TDEs are occasionally reported to be accompanied by a changing-look (CL) phenomenon in AGNs, wherein the appearance or disappearance of broad emission lines and thermal continuum from the accretion disk occurs (e.g., \citealt{eracleous_1995, merloni_2015, chan_2020, ricci_2020}). Although the origin of CL AGNs is unknown, the dramatic change in the accretion rate may be responsible for the type change (e.g., \citealt{penston_1984, elitzur_2014}). Indeed, fallback stellar debris can be a primary driver for enhancing accretion. For example, \citet{merloni_2015} and \citet{li_2022} argued that the width and central wavelength of the broad \hal\ emission changed substantially in some CL AGNs, possibly due to eccentric tidal debris originating from TDEs.

In this study, we report an exceptional IR-only flare occurring in NGC 3786, which may have originated from a TDE. This IR flare is highly obscured by circumnuclear dust and associated with the CL AGN phenomenon. The initial identification of the flare along with follow-up observations is described in Section 2. The physical properties derived from various observational datasets are summarized in Section 3. Finally, the physical origin of this flare is discussed in Section 4. Throughout the study, we assume a cosmology with $H_0=67.4$ km s$^{-1}$ Mpc$^{-1}$, $\Omega_m=0.315$, and $\Omega_{\lambda}=0.685$ \citep{planck_2020}. 

\section{Observation and Data}

\subsection{IR Flare in NGC 3786}
Initially, we used multi-epoch photometric data from WISE (\citealt{wright_2010}) to investigate the mid-IR variability of nearby galaxies within 50 Mpc. Because WISE observes certain targets multiple times during a few days in each visit, we calculate the representative magnitude in each visit as described in \citet{son_2022a} following the method of \citet{lyu_2019}. The $1\sigma$ uncertainty of the magnitude is defined as the root mean square sum of the measurement errors and the standard deviation of the measured brightnesses in the multiple observations in each visit. By comparing this dataset with light curves in the optical band, we serendipitously discovered an IR-only flare in NGC 3786, which appeared to have occurred around mid-2020. The amplitude of the bright IR flare was $\sim0.5$ and 0.8 mag in W1 and W2, respectively. These values are $6-8$ times larger than $\sigma$ of W1 and W2 before the flare. 

However, no significant increase in brightness was observed in the optical monitoring data from 2019-2021 obtained from the Zwicky Transient Facility (ZTF; \citealt{bellm_2019}; Fig. \ref{fig:fig1}). The RMS value of the $g$-band magnitude during 2020 is $\sim0.1$ mag, which is substantially less than the amplitude of the IR flare. This indicates that a corresponding flare is not observed in the optical band. As NGC 3786 has long been classified as an intermediate-type Seyfert (i.e., 1.8 or 1.9; \citealt{goodrich_1983}) with no detection of power-law continuum and broad \hb\ emission in the optical band, it is natural to suspect that an optical flare could be obscured by the dusty torus (e.g., \citealt{goodrich_1990}).
Note that a significant flux variation during $2014-2016$ may reveal that  another weak flare occurred. However, it is unclear if that event is directly associated with the IR flare in 2020. Interestingly, optical flares occasionally occurred in NGC 3786 over the past decades. (e.g, \citealt{nelson_1996a, koshida_2014}).

\subsection{Follow-up Observations}
To clarify the nature of the flare, we obtained optical and near-infrared (NIR) spectroscopic data using GMOS-N and GNIRS, respectively, at Gemini-North. Both spectra were taken on February 2, 2022, at an airmass of $1.1$ to $1.2$. The position angle was set to the parallactic angle to minimize light loss. The skies were clear and the seeing was measured as 0\farcs75-0\farcs8. For the optical data with GMOS-N, we used the B600 grating and a 0\farcs75 slit to cover \heii, \hb, and \hal\ simultaneously and achieve a spectral resolution of $\sim1100$ to properly measure the width of the emission lines. Two individual observation with an exposure time of 280 s were obtained via spectral dithering to fill the gap between the chips. The final optical spectrum covers a range of 4600-6861~\AA.

For the NIR spectrum with GNIRS, we adopted the cross-dispersed mode to obtain a wide spectral coverage from 0.8 to 2.5 $\mu$m. A 32~lines mm$^{-1}$ grating and 0\farcs3-wide slit were used to achieve a spectral resolution of 1800. From this instrument setup, a slit with a length of 7 arcsec is insufficiently long to obtain the sky spectrum for optimal sky subtraction. Therefore, we additionally acquired sky data in the blank field through target-sky-target nodding. 

Before the flare, the optical spectrum of NGC 3786 was obtained with the Perkins 1.8m telescope through a 2\asec-wide slit on April 6, 2010 (\citealt{koss_2017}). The spectrum covers $3900–7500$\AA\ with a spectral resolution of $\sim1050$. In addition, the pre-flare NIR spectrum obtained with the IRTF telescope is available, which was obtained on March 05, 2017. It was taken with SpeX Spectrograph and a ShortXD grating through a 0\farcs8-wide slit at an airmass of $1.2-1.4$. The spectrum covers from 0.7 to 2.55$\mu{\rm m}$ with a resolution of $\sim750$.

\subsection{Data Reduction}
 
The optical spectra were processed in a standard way, including bias subtraction, flat-field correction, and cosmic ray removal, using the \texttt{Gemini/GMOS} IRAF package. After performing wavelength calibration using arc images, we subtracted the sky emission sampled from the blank sky during the spectral extraction. The 1-d spectrum is extracted from an area of 0\farcs75$\times$5\asec\ centered at the nucleus. Finally, flux calibration was performed using the spectrum of Feige 66 (sdO).

The NIR spectrum was reduced using the \texttt{Gemini/GNIRS} IRAF package. For flat-fielding correction, we used two types of flat images obtained with different ramps to recover the response functions in low and high orders, simultaneously. Distortion correction was performed using pinhole spectra. The wavelength solution was determined using arc images obtained with the Argon lamp. After sky subtraction and spectral extraction with a diameter of 3 arcsec, we performed flux calibration and telluric correction using the standard star HIP 57239 (A2V) by adopting the method from \citet{vacca_2003}. The pre-flare NIR spectrum from the IRTF telescope was reduced in the same manner. Owing to a relatively large airmass difference ($\sim0.8$) between the standard star and the target galaxy in the IRTF observation, the telluric correction was imperfect, that highly degraded the data quality in the spectral region of \pal.

\begin{figure*}[t]
\centering
\includegraphics[width=0.95\textwidth]{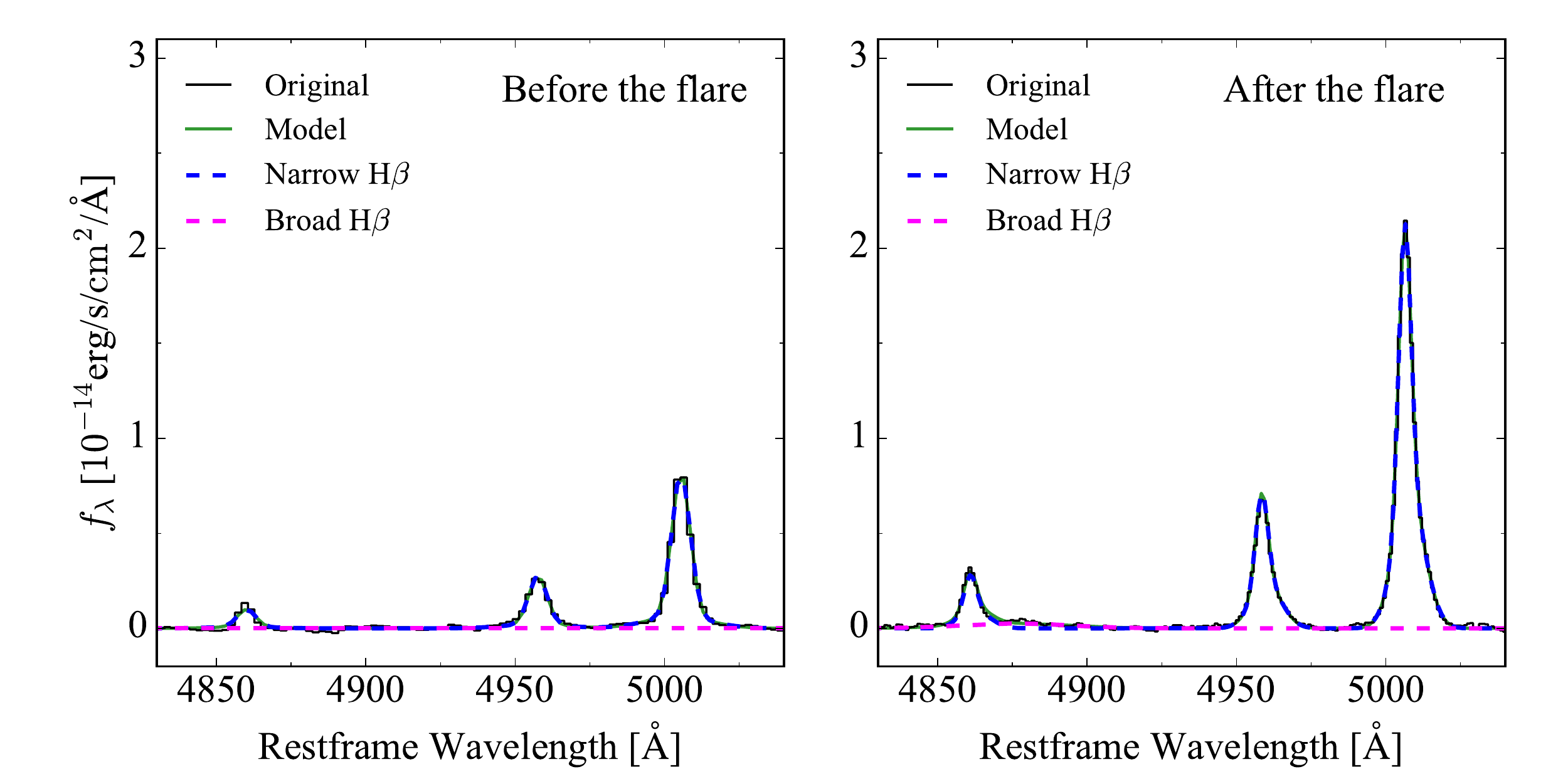}
\includegraphics[width=0.95\textwidth]{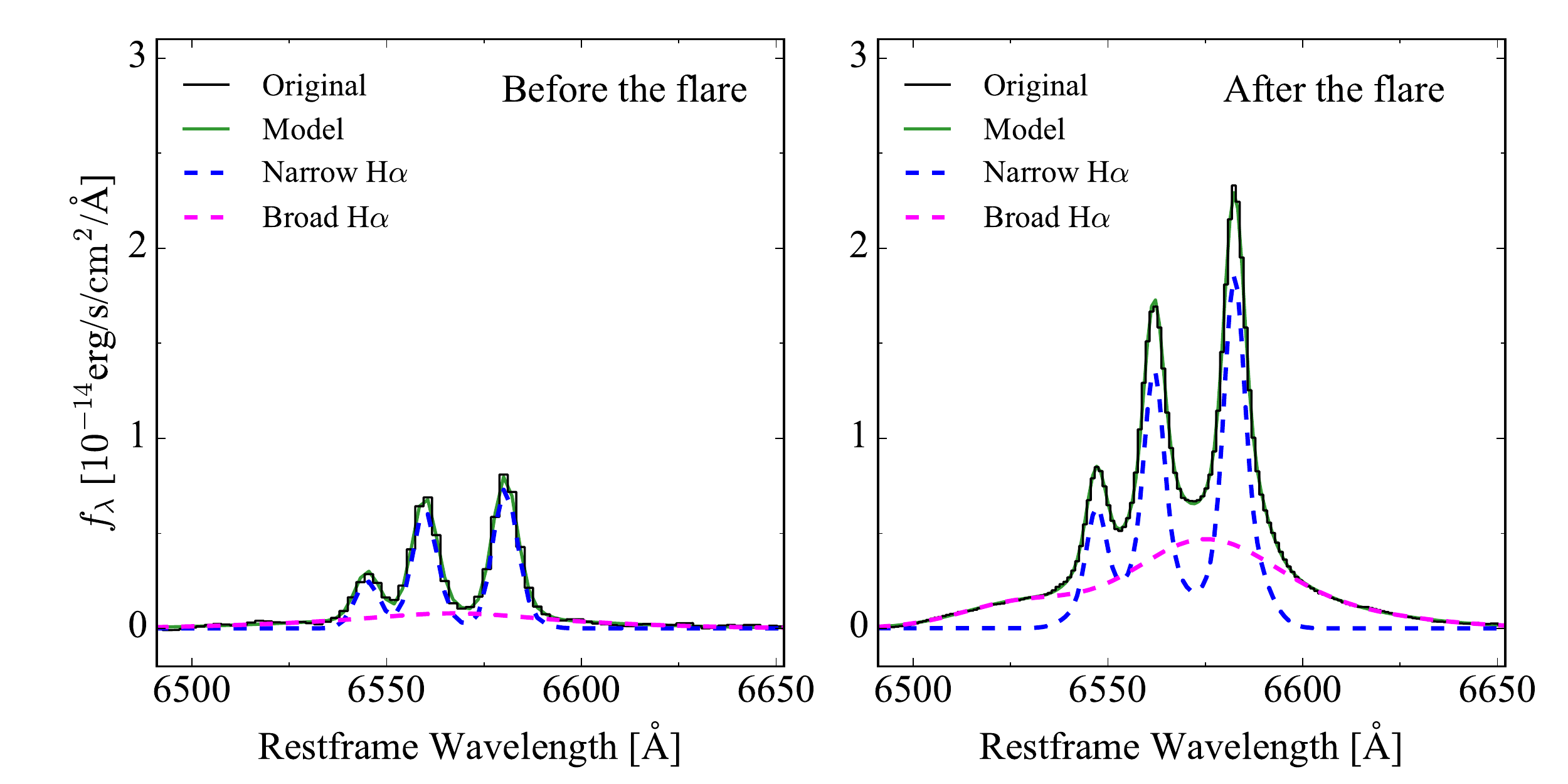}
\caption{
Spectra obtained with the Perkins 1.8m telescope before the flare (left) and with Gemini-North after the flare (right), for the \hb\ and \oiii\  region (top) and \hal\ and \nii\ region (bottom). In all panels, the histogram and green line represent the original spectra and best-fit model, respectively, whereas the blue and magenta dashed lines denote the best-fit models for the narrow and broad emission lines, respectively.
\label{fig:fig2}}
\end{figure*}

\begin{figure*}[t!]
\centering
\includegraphics[width=1.0\textwidth]{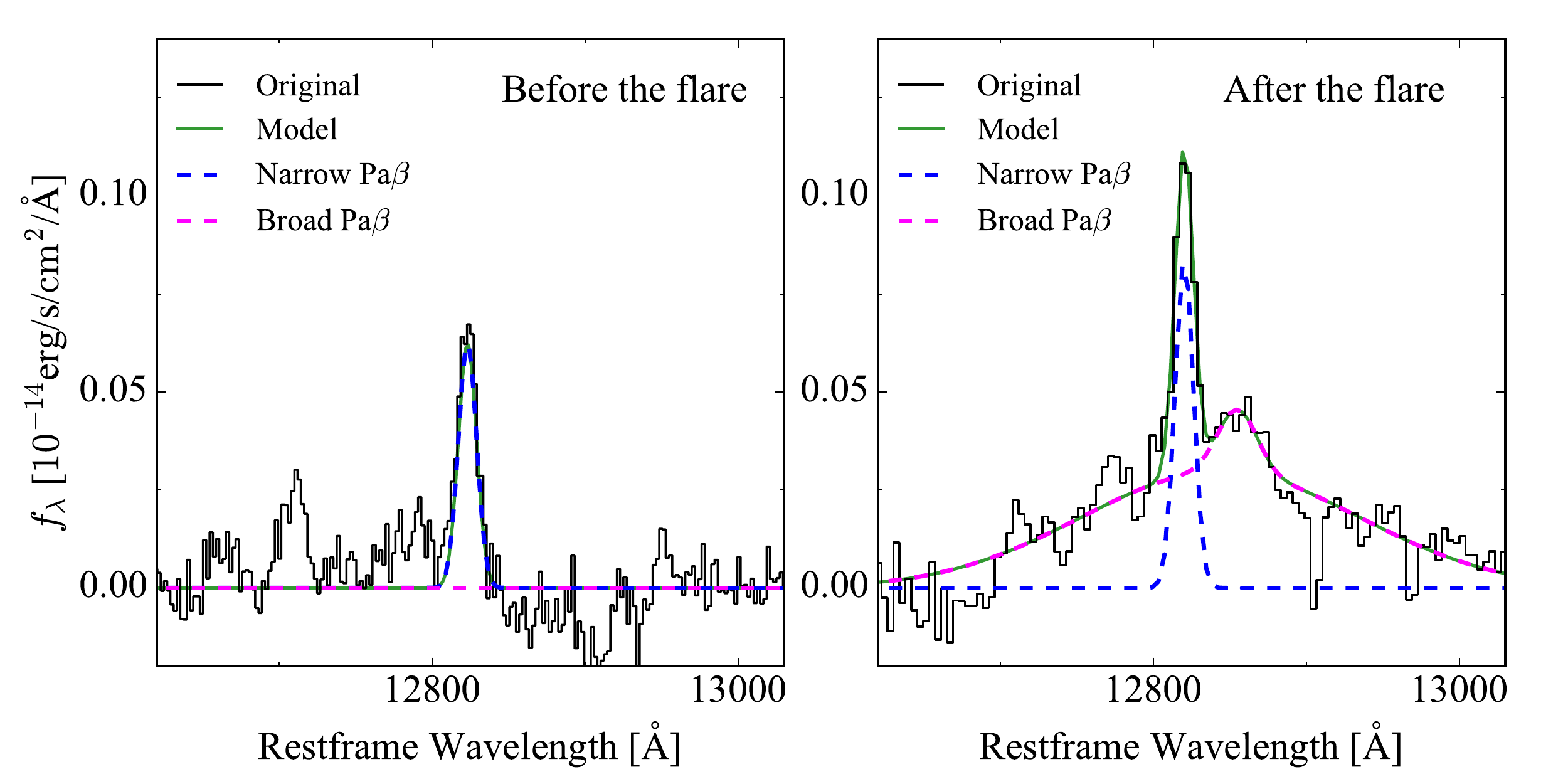}
\includegraphics[width=1.0\textwidth]{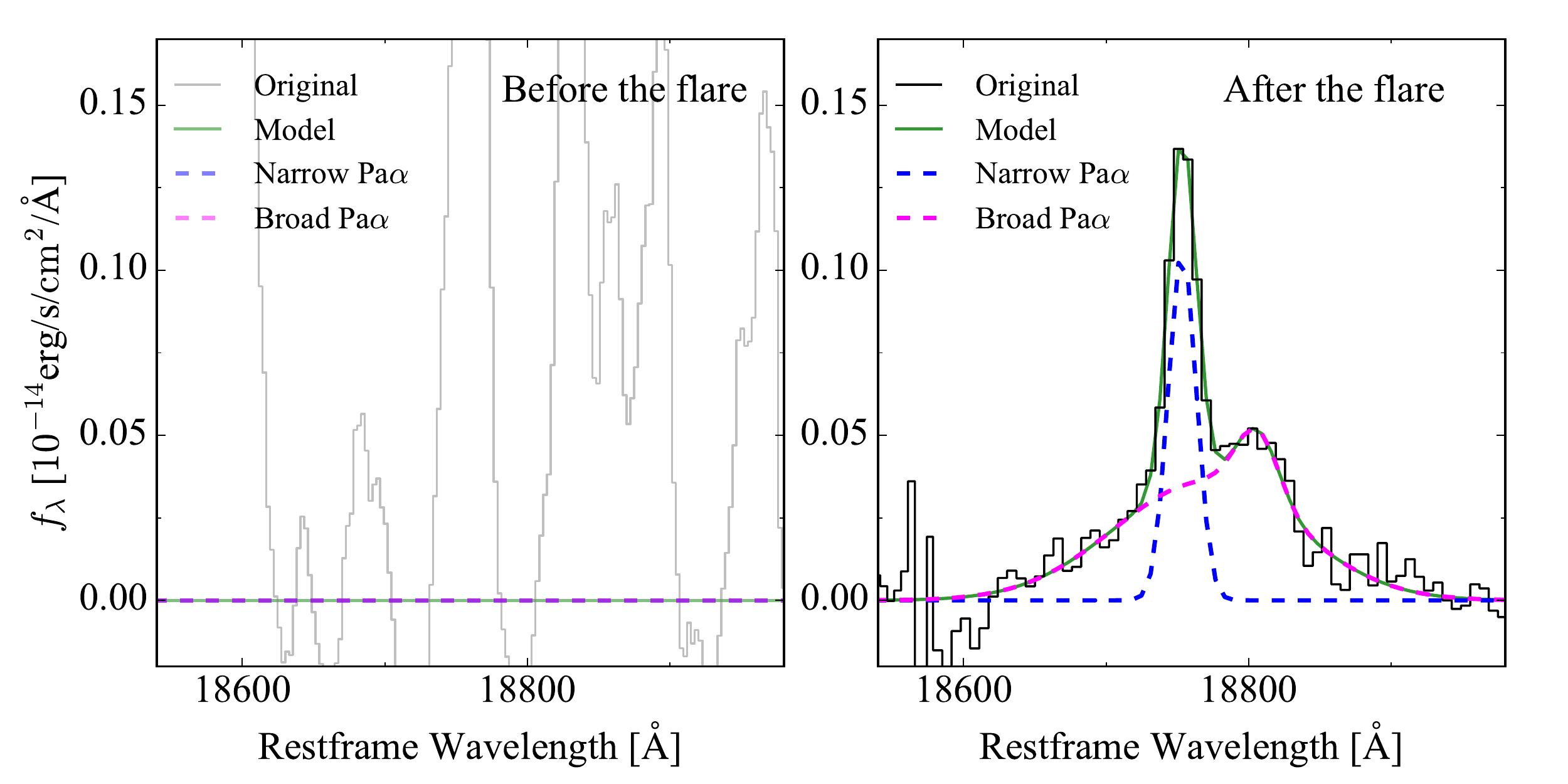}
\caption{
Fitting results for NIR spectra obtained before and after the IR flare. Spectra obtained with the IRTF before the flare (left) and with Gemini-North after the flare (right), for \pb\ (top) and \pal\ (bottom). Note that \pal\ data from IRTF are heavily compromised due to the imperfect telluric correction. Therefore, the emission-like feature around 18750\AA\ in the IRTF spectrum is an artifact caused by incomplete removal of the telluric features rather than \pal\ emission. 
\label{fig:fig3}}
\end{figure*}

\begin{figure*}[t!]
\centering
\includegraphics[width=1.0\textwidth]{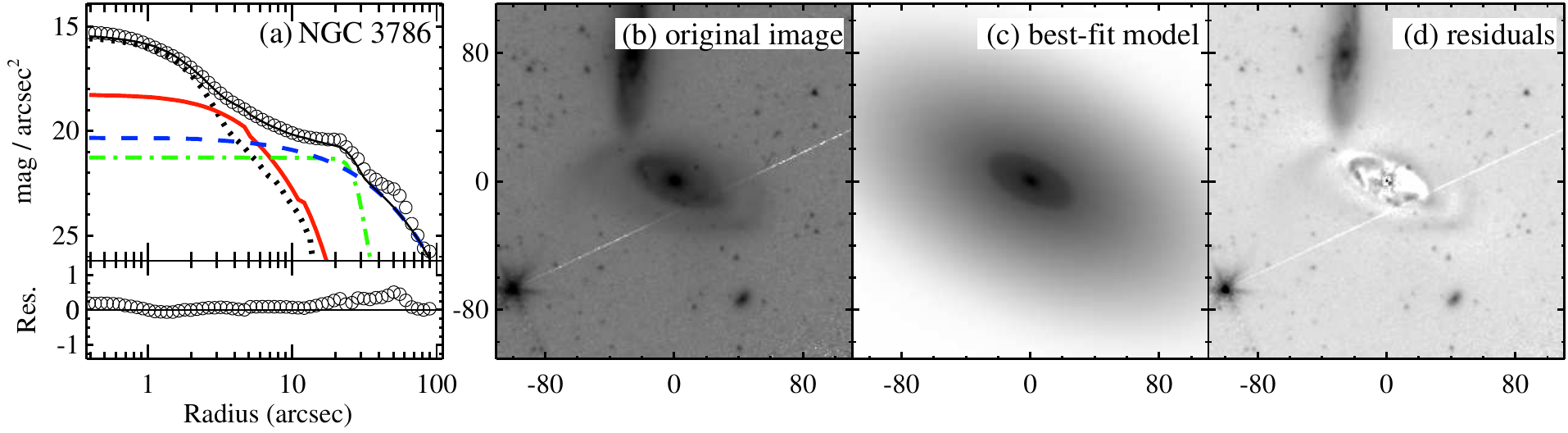}
\caption{
Results of 2D imaging decomposition of MIR imaging data obtained with Spitzer IRAC1 (3.6 $\mu$m). (a) Top panel shows surface brightness profiles of original data (open circles), nucleus (dotted line), bulge (red solid line), oval (green dotted-dashed line), and disk (blue dashed line) component. The residuals are displayed in the bottom panel. (b) Original image. (c) Best-fit model for only the host components. (d) Residuals.     
\label{fig:fig4}}
\end{figure*}

\section{Results}

\subsection{Spectral Fitting}
To compare the spectral properties before and after the flare, optical spectra obtained with a 1.8m Perkins telescope and GMOS-N were fitted using the following procedure. First, we simply modelled the local continuum using a first-order polynomial, which provides sufficiently reasonable results for the spectral measurements of the emission lines in AGNs (e.g., \citealt{denney_2009}). The \sii\ $\lambda\lambda$6716, 6731 doublet were fitted with four Gaussian components (i.e., two components for each line) to account for the putative outflow in the narrow line region (e.g., \citealt{kim_2006}) and used as a template profile to model other narrow lines (i.e., \nii, \hal). To fit the narrow emission lines for \oiii\ and \hb, we also used two Gaussian components for each line. The both narrow lines were assumed to be represented by the same profile. Where necessary, the broad lines of \hb\ and \hal\ were fitted with multiple Gaussian components. The optical spectrum obtained on April 6, 2010, exhibited a broad emission only in \hal. It is still possible that the weak broad \hb\ emission is buried by the noise in the continuum. We calculated the upper limit of the broad \hb\ flux ($\sim0.9\times10^{-14}$ erg/s/cm$^2$), using the $3\sigma$ of the underlying continuum and FWHM ($\sim 2695$ \kms) obtained from the post-flare spectrum. However, after the flare, the broad \hal\ emission was enhanced and a broad \hb\ emission appeared to be weakly present (Fig. \ref{fig:fig2}). More interestingly, the broad \hb\ emission was shifted toward longer wavelengths by $\sim900$ \kms. Furthermore, the broad \hal\ emission was also observed to be redshifted by $\sim600$ \kms\ with respect to the narrow emission line. Broad \heii\ $\lambda4686$ emission is often detected in the TDEs. However, we find no evidence of the broad \heii\ for both optical spectra. The upper limit of \heii\ is calculated using the 3$\sigma$ of the continuum and the FWHM ($\sim2695$ \kms) of the broad \hb\ emission in the GMOS spectrum, which yields the flux ratio of \heii\ to \hb\ is less than 0.92 in the GMOS spectrum. Fitting results are summarized in Table \ref{tab:table1}. 

We note that the narrow emission lines (e.g., \hb, \oiii, \hal, and \sii) in the post-flare spectrum appear to be enhanced compared to those from the pre-flare spectrum. It is unclear if this is due to either the intrinsic variation of the narrow line region or the difference in the observing conditions. The former is further discussed in \S{3.3}. One caveat is that the intensity from the extended region can be slightly overestimated due to the fact that the slit size is comparable and smaller than the seeing \citep{lee_2006}. As the flux calibration is performed using the standard star (i.e., the point source), the light loss owing to the small width of the slit can naturally lead to overestimating the flux of the extended emission in the science target. Therefore, the underlying continuum from the host stars and the emission from the narrow line region can be slightly overestimated while the fluxes of the broad emission lines originated from the unresolved nucleus is relatively free from this bias.

For the fitting of \pal\ and \pb\ in the NIR spectra for GNIRS, we used a fitting method similar to that used for the optical spectra, except that narrow emission lines were simply fitted with a single Gaussian as their shapes appear to be distinctive from the broad emission lines. From this analysis, we found that the broad emission lines were clearly detected but significantly redshifted with respect to the narrow lines by $\sim800$ \kms, which is in good agreement with the velocity shift in \hb. Using the NIR spectrum obtained with the IRTF telescope on March 5, 2017, we found that a broad component was not present, at least in \pb, before the flare. The upper limit of the broad \pb\ was calculated using the 3$\sigma$ of the continuum and FWHM measured from the post-flare NIR spectrum, that is significantly smaller than the flux of the broad \pb\ emission from the post-flare spectrum. It indicates that the broad \pal\ is likely to newly appear after the flare. Note that the IRTF spectrum around \pal\ was heavily affected by telluric absorption, which made it difficult to detect the broad component in \pal\ (Fig. \ref{fig:fig3}).

\subsection{2D Imaging Decomposition}
To estimate the BH mass of NGC 3786, we use the scaling relation between BH mass and bulge stellar mass. For this purpose, we conducted a 2D imaging decomposition of a 3.6 $\mu$m image obtained with the Spitzer Space Telescope Infrared Array Camera (IRAC) using \texttt{GALFIT} \citep{peng_2002}. The imaging data were obtained from the Spitzer Survey of Stellar Structure in Galaxies (S$^4$G; \citealt{sheth_2010}). The point spread function (PSF) adopted from \citet{salo_2015} was used for the convolution of the image and to model the tentative nucleus. Note that the PSF was properly rotated according to the orientation of the science image. The companion objects were masked using the masking image provided by \citet{salo_2015}. 

Based on the visual inspection, we attempted to model the host with three components (bulge, oval, and disk). While a range of \ser\ indices from 1 to 6 was used to fit the bulge component, we found that the bulge was represented best by a \ser\ with $n=1$. Finally, the host galaxy was fitted with three \ser\ components: one for the bulge with \ser\ index $n=1$, one for an oval with free $n$, and one for a disk with $n=1$ (Fig. \ref{fig:fig4}). In addition, to quantify the effect of the oval component on the bulge magnitude, we modelled the oval with the Ferrer function instead of the \ser\ component, which yielded results in terms of bulge brightness that were similar to within 0.04 mag. This result is consistent with the morphology (SABa) of NGC 3786 (\citealt{de_1991}). Because of the non-negligible amount of sky background, the sky value was simultaneously determined during the fitting procedure. The decomposition results are summarized in Table \ref{tab:table2}. 

The absolute magnitude of the bulge was estimated to be $-18.72$ mag. The typical uncertainty in the bulge luminosity is $\sim5\%$ (\citealt{salo_2015}). However, if the PSF mismatch is severe, the uncertainty can be increased up to $\sim30\%$ (e.g., \citealt{kim_2008a,kim_2017,son_2022b}). To compute the stellar mass of the bulge, we used the mass-to-light ratio given by \citet{munoz_2013}. It yields $M_{*,{\rm bul}}\approx10^{9.62}M_\odot$, while the total stellar mass of the host is $\sim10^{10.66}M_\odot$. The bulge-to-total light ratio ($B/T$) is $\sim0.09$.

\subsection{AGN Properties}
As the physical properties of the target galaxy are essential for investigating the physical origin of the flare, we estimated BH masses, bolometric luminosities, and Eddington ratios using various methods. For a type 1 AGN, BH masses can be computed using the virial method from the equation $M_{\rm BH} \sim \frac{v^2 R}{G}$, where $R$ and $v$ are the radius and velocity dispersion of the broad line region (BLR), respectively. From the empirical relation between AGN luminosity and $R$, the BH mass can be inferred using the single-epoch spectrum (e.g., \citealt{bentz_2013}). Note that the broad-line luminosity instead of the optical continuum luminosity to estimate the BLR radius, which is needed to calculate the BH mass, because the AGN continuum is not detected in the spectrum. Note that the BH mass from the virial method is associated with an uncertainty, namely a scaling factor ($f$), which may depend on the geometry and kinematics of the BLR.  

Conversely, BH mass can be measured independently by adopting the tight correlation between the BH mass and bulge mass (\mm). Therefore, we estimated the BH mass before the flare using the two measurements: using the $3.6\mu$m bulge luminosity (\S{3.2}) and the broad \hal\ emission. It is known that \mm\ relation as well as the scaling factor in the virial method depends on the bulge type of the host galaxies (e.g., \citealt{kormendy_2013, ho_2014}. For the purpose of the consistency, we adopted the \mm\ relation and the virial estimates derived from all galaxies regardless of the bulge type \citep{kormendy_2013, greene_2005, ho_2015}. We obtained the value of $\log M_{\rm BH}/M_\odot = 6.70$ and $6.68$, respectively. The BH mass from the virial method is consistent with that from the \mm\ relation, despite the fact that the broad line region can be moderately obscured.  Additionally, we also used the correlation between the BH mass and stellar velocity dispersion (\msig\ relation) to derive the BH mass. We used two measurements of the stellar velocity dispersion. \citet{nelson_1995} reported $\sigma_*\sim 142\pm13$\kms, based on the spectral fitting around Ca II triplet absorption. We also utilized the pre-flare optical spectrum to independently estimate the stellar velocity dispersion as it covers Ca H+K stellar absorption, which helps to robustly constrain the kinematics of the stars. By applying the Penalized PiXel-Fitting (pPXF) method \citep{cappellari_2004} to the spectrum, we found $\sigma_*\sim 105\pm59$\kms. As a result, the black hole masses inferred from the \msig\ relation of all types of bulges from \citet{kormendy_2013} are $\log M_{\rm BH}/M_\odot = 6.83$ and $7.54$.


We applied the same methods to calculate BH mass after the flare, wherein broad emissions from NIR spectrum were additionally used. From the virial methods, BH masses were estimated as $\log M_{\rm BH}/M_\odot = 6.82$, 6.76, and 7.23 from the \hal, \pal, and \pb\ emissions, respectively \citep{kim_2010} without the extinction correction. By comparing the observed line ratios among \hal, \pal, and \pb\ with the intrinsic values from \citet{kim_2010}, we found that $E(B-V)$ ranges from 0 to 0.51 (\citealt{kimd_2018b}). We again calculated BH masses using the extinction corrected luminosity by adopting the color excess ($E(B-V)\sim0.35$) derived from the flux ratio of \pal\ to \hal, yielding a slight increase in the BH masses ($\log M_{\rm BH}/M_\odot = 6.97$, 6.79, and 7.28 from the \hal, \pal, and \pb\ emissions, respectively).    


Although those estimates are in good agreement with that derived from the bulge mass within the uncertainty ($0.4-0.5$ dex in the virial methods; \citealt{park_2012, shen_2013}), the BH masses estimated after the flare appears to be systematically larger than those estimated before the flare. If the newly appeared broad emission lines originated from the TDE, it is possible that the gas in the BLR is not virialized, which can introduce a systematic bias in the BH mass estimation (e.g., \citealt{li_2022}). Therefore, we use $\log M_{\rm BH}/M_\odot = 6.70$, derived from the bulge mass, as the best estimate because of the unknown bias in the virial estimates. Nevertheless, the majority of the BH mass measurements ($\log M_{\rm BH}/M_\odot \sim 6.5 - 7.5 $ from various methods are in broad agreement within the typical uncertainties ($0.3-0.5$ dex).  

The bolometric luminosity ($L_{\rm bol}$) can be inferred either from monochromatic luminosities at various bands or line luminosities with the proper bolometric conversion. For a type 2 or intermediate-type AGN, the \oiii\ emission line is often used as an estimator of bolometric luminosity because the featureless continuum and broad emission are not detected. The conversion from the \oiii\ luminosity to the bolometric luminosity can be carried out in two different ways: (1) using the conversion factor computed from the observed \oiii\ luminosity not corrected for the extinction (e.g., \citealt{heckman_2004}); (2) using the conversion factor derived from the extinction-corrected \oiii\ luminosity (e.g.,  \citealt{lamastra_2009}).
By adopting the conversion factor from \citet{heckman_2004} along with the \oiii\ luminosity ($L_{\rm [O \,III]}=10^{40.50}$ \lum) after the flare, the bolometric luminosity was estimated to be $10^{44.04}$ \lum. From the extinction-corrected \oiii\ luminosity with the Balmer decrement ($f_{{\rm H}\alpha}/f_{{\rm H}\beta}\approx5.6$), we found $L_{\rm bol}=10^{43.44}$ \lum\ by adopting the conversion factor from \citet{lamastra_2009}.

Note that the \oiii\ luminosity appeared to be enhanced after the flare. However, the observing conditions (e.g., slit width, position angle, and extraction aperture) in both observations, which can easily affect the flux measurements in the extended narrow line region, are not identical each other. Additionally, \oiii\ fluxes measured from past observations carried out during the last $\sim40$ years appeared to vary by a factor of $\sim5$ (e.g., \citealt{goodrich_1983, keel_1985, dahari_1988, cruz-gonzalez_1994, koss_2017}), although it is again difficult to determine the origin of this flux variation as the observing conditions are not the same. Narrow-line flux is generally believed to be invariant on years timescale, but the flux variations in \oiii\ emission have been reported in CL AGNs and TDEs (e.g., \citealt{denney_2014, barth_2015, li_2022}). Previous studies have also shown that NGC 3786 has undergone extreme flux variations (e.g., \citealt{nelson_1996a, koshida_2014}) and type transition from type 1.8 (e.g., \citealt{goodrich_1983, osterbrock_1993}) to 1.9 (e.g., \citealt{trippe_2010, koss_2017}) over the past decades. Moreover, NGC 5548, one of the nearest AGN, exhibits the flux variation of \oiii\ emission with a time scale of a few years possibly due to the fact that the ionized gas in NLR is highly concentrated in the center and the gas density is significantly higher than expected so that the recombination time scale becomes smaller than a year \citep{peterson_2013}. Therefore, the genuine change of the narrow line region could also be responsible for \oiii\ flux variation of NGC 3786. In any case, the \oiii\ luminosity is not suitable for an immediate trace of the bolometric luminosity due to its uncertainty in the measurements and a relatively long response time scale with respect to the variation of the AGN continuum.

Alternatively, we used W1 mag for the estimation of $L_{\rm bol}$ using the conversion equation in \citet{son_2022a}. To properly remove the host contribution in the W1 band, we performed spectral energy density fitting using the template spectra of host galaxies and AGNs \citep{son_2022a}. From this experiment, we determined that $L_{\rm bol}=10^{42.89}$ \lum\ before the flare. Alternatively, the MIR magnitude of the nucleus can be simply estimated from the imaging decomposition of the {\it Spitzer} data (\S{3.2}). The nuclear magnitude ($\sim 13.2$ mag) from this method is significantly smaller (brighter) than that ($\sim 14.7$ mag) from the SED fitting, yielding that $L_{\rm bol}=10^{43.46}$ \lum. It is uncertain what causes this discrepancy. But the PSF mismatch can severely introduce systematic uncertainties not only for the bulge magnitude but also for the nuclear magnitude. Moreover, the {\it Spitzer} data was obtained on December 17, 2004, while the WISE data was taken after 2010. Therefore, the long-term variation in MIR brightness can be also responsible for this discrepancy.

Finally, from the luminosity of the \hal\ emission before the flare, $L_{\rm bol}=10^{42.66}$ \lum\ \cite[]{greene_2005, richards_2006}. After the flare, by applying the same methods, $L_{\rm bol}=10^{43.27}$ and $10^{43.35}$ \lum\ were estimated from \hal\ and W1 mag, respectively. Additionally, we used the \pal\ luminosity to compute the bolometric luminosity ($L_{\rm bol}=10^{43.52}$ \lum) by adopting the conversion factor from \citet{kim_2022}. Based on the above measurements, the Eddington ratio was calculated under the assumption that $\log M_{\rm BH}/M_\odot = 6.70$ (Tab. \ref{tab:table3}). 

\section{Physical Origins of the Flare}

Here, we discuss the physical origin of the MIR-only flare in NGC 3786. Variations in MIR color can be used as a probe for flares. W1$-$W2 became redder as it brightened in the W1-band (Fig. \ref{fig:fig1}), which is in broad agreement with canonical CL AGNs and low-luminosity AGNs (e.g., \citealt{yang_2018, yang_2019, son_2022a}), and distinctive from the color variability in supernovae (SN). Therefore, the flare is likely to originate from nuclear activity rather than SN.    

An intriguing result from the follow-up observations was that the newly appearing broad emission lines were redshifted up to 900 \kms. Such a high-velocity offset is often observed in CL AGNs, possibly resulting from the eccentric tidal debris generated from the TDE phenomenon (e.g., \citealt{merloni_2015, li_2022}). In this light, it is natural to suspect that the MIR flare was caused by radiation of circumnuclear dust, which is heated by the enhanced light from the accretion disk due to the TDE. 

However, \heii\ $\lambda4686$, which is most commonly found in ordinary TDEs (e.g., \citealt{arcavi_2014}), was not detected in the GMOS spectrum obtained after the flare. In general, \heii\ is known to be more luminous than \hb\ in TDEs. However, the flux ratio between the two lines and its variation over time substantially varies among TDEs (e.g., \citealt{hung_2017}). Given that the follow-up observation was performed at least $\sim2$ years after the flare, it may not be surprising that \heii\ dimmed and is not detected. Alternatively, it is also possible that the gas density is not yet sufficiently low for He II to emerge after \hal\ enhancement (\citealt{li_2022}). In addition, dust obscuration may not be negligible, which can lead to significant attenuation of \heii\ flux.  

It is well-known that TDE preferentially occurs in low-mass BHs ($\leq 10^{7.5}M_\odot$) because at high-mass BHs ($> 10^{7.5}M_\odot$) the tidal radius of low-mass stars becomes less than the Schwarzschild radius (e.g., \citealt{stone_2016}). NGC 3786 ($M_{\rm BH}\sim10^{6.70}M_\odot$) appears to meet the BH mass range for a high TDE occurrence rate, which implies that the TDE scenario is still favorable. Note that the BH masses derived from the \msig\ relation are slightly larger ($M_{\rm BH}\sim10^{6.83-7.54}M_\odot$). However, even with these values we still cannot completely rule out the TDE scenario, accounting for the intrinsic scatter of the \msig\ relation ($\sim0.3$ dex). In addition, at a given stellar velocity dispersion, galaxies with pseudo-bulges systematically have a smaller BH mass compared to ellipticals or galaxies with classical bulges. Our imaging decomposition results indicate that the NGC 3786 can be well fit with a pseudo-bulge represented by the small \ser\ index and $B/T$. Therefore, BH mass inferred from the \msig\ relation can be somehow overestimated.

In contrast, the CL AGN phenomenon can originate from the transition of the accretion mode (e.g., \citealt{guolo_2021}). For example, in the AGN with the low Eddington ratio, the BLR can physically disappear in the disk-wind driven BLR model due to the low efficiency in the accretion disk or lack of ionizing photon (e.g., \citealt{elitzur_2009, elitzur_2014}). This transition between the standard thin accretion disk model and a radiatively inefficient accretion flow (RIAF) can occur at the critical Eddington ratio ($\sim1\%$; e.g., \citealt{ho_2008}). Interestingly, the Eddington ratio of the nucleus of NGC 3786 before and after the flare ($\sim0.7-3.5\%$) is in broad agreement with the critical value. Even if we adopt the BH masses derived from \msig\ relation, the estimated Eddington ratios range from $\sim0.1\%$ to $0.5\%$, which is still comparable to the critical value for the transition. This suggests that the flare and the CL AGN phenomenon can be driven simply by accretion mode transition, although the velocity offset of the broad emission lines may not be naturally explained by this mechanism.  


In summary, while the increase of the accretion rate cannot be entirely excluded as the physical origin of the IR flare, the TDE is more favorable to explain the velocity shifts in the broad emission lines (e.g., \citealt{merloni_2015, li_2022}). In this scenario, the redshifted broad emission lines are likely to arise from the tidal debris on eccentric orbits (e.g., \citealt{guillochon_2014}). To further determine the nature of the flare, multi-epoch spectroscopic data will be essential because the BLR can be shifted on a scale of $\sim$ month if the BLR is generated from tidal debris (e.g., \citealt{zabludoff_2021, li_2022}). For the systematic studies of the heavily obscured flares, the future IR multi-epoch survey, such as SPHEREx, will play a crucial role by extensively detecting MIT flares in the galactic nuclei (e.g., \citealt{dore_2018, kim_2021b}). 

\begin{acknowledgments}
We thank an anonymous referee for her/his constructive comments that helped to improve the manuscript. LCH was supported by the National Science Foundation of China (11721303, 11991052, 12011540375) and the China Manned Space Project (CMS-CSST-2021-A04, CMS-CSST-2021-A06). This work was supported by the National Research Foundation of Korea (NRF) grants (No.\ 2020R1A2C4001753 and No.\ 2022R1A4A3031306) funded by the Korean government  (MSIT) and under the framework of international cooperation program managed by the National Research Foundation of Korea (NRF-2020K2A9A2A06026245). DK was supported by the National Research Foundation of Korea (NRF) grant funded by the Korea government (MSIT) (No. 2021R1C1C1013580). This work was supported by K-GMT Science Program (PID: GN-2022A-FT-203) of Korea Astronomy and Space Science Institute (KASI). Based on observations obtained at the international Gemini Observatory, a program of NSF’s NOIRLab, which is managed by the Association of Universities for Research in Astronomy (AURA) under a cooperative agreement with the National Science Foundation. on behalf of the Gemini Observatory partnership: the National Science Foundation (United States), National Research Council (Canada), Agencia Nacional de Investigaci\'{o}n y Desarrollo (Chile), Ministerio de Ciencia, Tecnolog\'{i}a e Innovaci\'{o}n (Argentina), Minist\'{e}rio da Ci\^{e}ncia, Tecnologia, Inova\c{c}\~{o}es e Comunica\c{c}\~{o}es (Brazil), and Korea Astronomy and Space Science Institute (Republic of Korea).

\end{acknowledgments}

\begin{deluxetable*}{lcrrrr}
\tablecolumns{13}
\tablenum{1}
\tablewidth{0pc}
\tablecaption{Spectral Properties of NGC 3786 \label{tab:table1}}
\tablehead{
\colhead{\h Line} &
\colhead{\h Date} &
\colhead{\h Flux} &
\colhead{\h FWHM} &
\colhead{\h Center} &
\colhead{\h Velocity shift} \\
\colhead{\h (1)} &
\colhead{\h (2)} &
\colhead{\h (3)} &
\colhead{\h (4)} &
\colhead{\h (5)} &
\colhead{\h (6)}
}
\startdata
Broad \heii &\h b &\h$\leq0.8$    &\h \nd &\h \nd \\  
         \h &\h a &\h$\leq1.1$    &\h \nd &\h \nd \\ 
\hline
Narrow \hb  &\h b &\h 1.1   &\h 383 &\h 4859.4\\  
         \h &\h a &\h 2.1   &\h 243 &\h 4860.8\\ 
\hline
Broad \hb   &\h b   &\h$\leq0.9^{\rm a}$ &\h \nd   &\h \nd     \\
         \h &\h a$^{\rm b}$           &\h 1.2   &\h 2695  &\h 4875.8 &\h +928\\ 
\hline
\oiii$\lambda5007$&\h b &\h 7.5   &\h 383 &\h 5006.4\\
         \h       &\h a &\h 16.2  &\h 243 &\h 5006.6\\ 
\hline
Narrow \hal &\h b &\h 5.2   &\h 224 &\h 6560.6\\  
         \h &\h a &\h 11.7   &\h 174 &\h 6562.2\\ 
\hline
Broad \hal  &\h b           &\h 5.8   &\h 2854 &\h 6567.1 &\h +297\\  
         \h &\h b$^{\rm s}$ &\h 1.0   &\h 1647 &\h 6567.7 &\h +324\\ 
         \h &\h b$^{\rm s}$ &\h 4.8   &\h 3938 &\h 6565.5 &\h +224\\ 
         \h &\h a           &\h 30.2  &\h 2303 &\h 6575.3 &\h +598\\ 
         \h &\h a$^{\rm s}$ &\h 14.0  &\h 1831 &\h 6574.3 &\h +553\\ 
         \h &\h a$^{\rm s}$ &\h 12.0  &\h 3421 &\h 6583.5 &\h +973\\ 
         \h &\h a$^{\rm s}$ &\h  4.2  &\h 1571 &\h 6526.9 &\h $-1613$\\ 
\hline
\sii$\lambda\lambda6717,6731$ &\h b &\h 4.5   &\h $224^{\rm c}$ &\h $6713.3^{\rm c}$\\
         \h                   &\h a &\h 9.4   &\h $174^{\rm c}$ &\h $6715.4^{\rm c}$\\ 
\hline
Narrow \pb  &\h b &\h 0.94  &\h $\cdots^{\rm d}$ &\h 12823.9\\  
        \h  &\h a &\h 1.27  &\h 298 &\h 12819.0\\ 
\hline
Broad \pb   &\h b &\h$\leq4.70^{\rm a}$  &\h \nd  &\h \nd \\  
        \h  &\h a &\h 7.30  &\h 3108 &\h 12853.9 &\h +816\\ 
        \h  &\h a$^{\rm s}$ &\h 0.52  &\h 688 &\h 12854.9 &\h +840\\ 
        \h  &\h a$^{\rm s}$ &\h 6.78  &\h 5025 &\h 12843.8 &\h +580\\ 
\hline
Narrow \pal &\h b &\h \nd  &\h \nd &\h \nd \\  
         \h &\h a &\h 2.60  &\h 348 &\h 18750.7\\ 
\hline
Broad \pal  &\h b &\h \nd &\h \nd &\h \nd \\
         \h &\h a &\h 6.54  &\h 1831 &\h 18802.8 &\h +834 \\
         \h &\h a$^{\rm s}$ &\h 0.81  &\h 543 &\h 18806.3 &\h +890 \\
         \h &\h a$^{\rm s}$ &\h 5.73  &\h 2398 &\h 18769.8 &\h +306 \\
\enddata
\tablecomments{
Col. (1): Emission line. 
Col. (2): Observation epoch: ``b''=before the flare, ``a''=after the flare.
Col. (3): Line flux in units of $10^{-14}$ erg/s/cm$^2$.
Col. (4): FWHM of the emission line in units of \kms.
Col. (5): Restframe central wavelength of the emission line in units of \AA.
Col. (6): Velocity shifts of the broad emission lines relative to the narrow emission lines in units of \kms. \\
$^{\rm a}$ Upper limit of the emission line derived from $3\sigma$ of the underlying continuum and FWHM measured from the spectra taken after the flare.  \\
$^{\rm b}$ Single Gaussian component is used to fit the broad \hb. \\ 
$^{\rm c}$ Measurements for \sii$\lambda6717$. \\ 
$^{\rm d}$ The line is not resolved due to the low spectral resolution of the IRTF spectrum ($R\sim750$). \\
$^{\rm s}$ Measurements for each Gaussian component.
}
\end{deluxetable*}

\begin{deluxetable*}{clrr}
\tablecolumns{13}
\tablenum{2}
\tablewidth{0pc}
\tablecaption{Host Properties of NGC 3786 \label{tab:table2}}
\tablehead{
\colhead{\h Component} &
\colhead{\h $m_{3.6\mu{\rm m}}$} &
\colhead{\h $n$} &
\colhead{\h $R_e$} \\
\colhead{\h (1)} &
\colhead{\h (2)} &
\colhead{\h (3)} &
\colhead{\h (4)} 
}
\startdata
Nucleus &\h $13.22$ &\h \nd  &\h \nd \\  
Bulge  &\h $14.30$ &\h 1.00 &\h 3.42\\
Oval &\h $13.68$ &\h 0.08 &\h 19.29\\
Disk  &\h $12.43$ &\h 1.00 &\h 32.37
\enddata
\tablecomments{
Col. (1): Component. 
Col. (2): AB magnitude in the IRAC1 band.
Col. (3): \ser\ index.
Col. (4): Effective Radius in the unit of arcsec.
}
\end{deluxetable*}

\begin{deluxetable*}{clrr}
\tablecolumns{13}
\tablenum{3}
\tablewidth{0pc}
\tablecaption{Properties of NGC 3786 \label{tab:table3}}
\tablehead{
\colhead{\h Property} &
\colhead{\h Data} &
\colhead{\h Before the flare} &
\colhead{\h After the flare} \\
\colhead{\h (1)} &
\colhead{\h (2)} &
\colhead{\h (3)} &
\colhead{\h (4)} 
}
\startdata
$\log M_{\rm BH}$ &\h $L_{3.6\mu \rm m, bulge}$ &\h 6.70 &\h $\cdots$\\
\h &\h $\sigma_*$ &\h $6.83-7.54$ &\h \\
\h &\h $L_{{\rm H}\alpha}$ &\h 6.68 &\h 6.82\\
\h &\h $L_{{\rm Pa}\alpha}$ &\h $\cdots$ &\h 6.76\\
\h &\h $L_{{\rm Pa}\beta}$  &\h $\cdots$ &\h 7.23\\
\hline
$\log L_{\rm bol}$  &\h $L_{[\rm O\, III]}$ &\h 43.70 &\h 44.04\\
\h  &\h $L_{[\rm O\, III]}^{\rm a}$  &\h 43.14 &\h 43.44\\
\h  &\h $L_{{\rm H}\alpha}$  &\h 42.66 &\h 43.27\\
\h  &\h $L_{{\rm Pa}\alpha}$  &\h $\cdots$ &\h 43.52\\
\h  &\h $M_{\rm W1}^{\rm a}$ &\h 42.88 &\h 43.35 \\
\h  &\h $M_{\rm W1}^{\rm b}$ &\h 43.47 &\h $\cdots$ \\
\hline
$\log (L_{\rm bol}/L_{\rm Edd})$ &\h $L_{[\rm O\, III]}$  &\h $-1.10$ &\h $-0.76$ \\
\h  &\h $L_{[\rm O\, III]}^{\rm c}$ &\h $-1.66$ &\h $-1.36$ \\
\h  &\h $L_{{\rm H}\alpha}$  &\h $-2.14$ &\h $-1.53$ \\
\h  &\h $L_{{\rm Pa}\alpha}$ &\h $\cdots$ &\h $-1.28$ \\
\h  &\h $M_{\rm W1}^{\rm a}$ &\h $-1.92$ &\h $-1.45$ \\
\h  &\h $M_{\rm W1}^{\rm b}$ &\h $-1.33$ &\h $\cdots$
\enddata
\tablecomments{
Col. (1): AGN properties. 
Col. (2): Observational data used to estimate the AGN property.
Col. (3): Estimates based on the data obtained before the flare.
Col. (4): Estimates based on the data obtained after the flare.\\
$^{\rm a}$ W1 magnitude derived from the SED fit. \\
$^{\rm b}$ W1 magnitude derived from the imaging decomposition of the Spitzer 3.6$\mu{\rm m}$ data taken in 2004. \\
$^{\rm c}$ Extinction-corrected.
}
\end{deluxetable*}

\bibliography{torus}

\end{document}